\documentclass{ws}
\newcommand{\ind}[2]{^{#1}_{\mbox{\scriptsize #2}}}
\newcommand{\al}[2]{\alpha\ind{#1}{#2}}
\newcommand{\hal}[2]{\widehat{\alpha}\ind{#1}{#2}}
\newcommand{\ro}[1]{\rho^{(#1)}}
\newcommand{\Ro}[2]{{\cal R}^{(#1)}(#2)}
\newlength{\intwidth}
\newcommand{\pvint}[2]{
\settowidth{\intwidth}{$\displaystyle\int$}
\mbox{\hbox to 0pt{\hbox
to\intwidth{{\hfill{\raisebox{0.45mm}{\scriptsize$-$}}\hfill}}\hss}%
$\displaystyle\int_{#1}^{#2}$}}
\def\fpb{\frac{4 \pi}{\bz}}
\def\Nc{N_{\mbox{\scriptsize c}}}
\def\nf{n_{\mbox{\scriptsize f}}}
\def\bz{\beta_0}
\def\mpi{m_{\pi}}
\def\KL{K\"all\'en--Lehmann }
\def\epem{e^{+}e^{-}}

\begin{document}
\markboth{}{}
\catchline{}{}{}{}{}

\title{THE QCD ANALYTIC EFFECTIVE CHARGE \\
AND ITS DEPENDENCE ON THE PION MASS}

\author{A.V.~NESTERENKO$^1$ and J.~PAPAVASSILIOU$^2$\vskip2.5mm}

\address{$^1$Bogoliubov Laboratory of Theoretical Physics,
Joint Institute for Nuclear Research \\
Dubna, 141980, Russian Federation\vskip-2.5mm}

\address{$^2$Departamento de F\'\i sica Te\'orica and IFIC,
Centro Mixto,\\ Universidad de Valencia-CSIC,
E-46100, Burjassot, Valencia, Spain}

\maketitle

\pub{Received 18 September 2004}{}

\begin{abstract}
A new model for the QCD analytic running coupling, which incorporates the
effects due to the $\pi$~meson mass, is proposed. The properties of this
invariant charge in spacelike and timelike regions are examined. Its main
distinctive features are a finite infrared limiting value, which depends
on the pion mass, and the ``plateau--like'' behavior in the deep infrared
domain of the timelike region.
\end{abstract}

\section{Introduction}

     The description of the infrared hadron dynamics remains a crucial
challenge for elementary particle physics. Whereas perturbative
calculations allow for a detailed study of the strong interaction
processes at high energies, to date there is no reliable theoretical
method which would enable one to handle the hadron dynamics at low
energies. However, many physical phenomena are tightly bound to the
intrinsically nonperturbative aspects of the strong interaction.

     The renormalization group (RG) method plays a fundamental role in the
framework of Quantum Field Theory (QFT) and its applications. Usually, in
order to describe Quantum Chromodynamics (QCD) in the ultraviolet region,
one applies the RG method together with perturbative calculations.
Eventually, this leads to approximate solutions of the RG equations, which
are commonly used in the quantitative analysis of high energy processes.
However, such solutions possess unphysical singularities in the infrared
domain, a fact that contradicts the general principles of local QFT and
complicates the theoretical description of the intermediate and low energy
experimental data. Nevertheless, these difficulties can be avoided if one
complements the perturbative results with a proper nonperturbative
insight.

     For this purpose the relevant dispersion relations can be used. The
idea of employing the information ``stored'' in such integral
representations together with perturbation theory forms the basis of the
analytic approach to~QFT. First it was proposed in the framework of
Quantum Electrodynamics\cite{AQED} and lately it was extended\cite{ShSol}
to~QCD. Another important application of the dispersion relations is the
analysis of the timelike QCD experimental data. Here, the dispersion
relation for the Adler $D$ function\cite{Adler} enables one to handle the
timelike strong interaction processes in a consistent way.

     The effects due to the masses of the lightest hadrons can be safely
neglected only when one studies the QCD processes at high energies. But
for the description of the low energy dynamics the mass effects become
essential. For example, this is crucial for the description of the
inclusive $\tau$~lepton decay. Both, the perturbative results and the
dispersion relation for the Adler $D$ function with the nonvanishing pion
mass, are essential here for the correct interpretation of these data.
However, no such mass effects have been taken into account in the analytic
approach to QCD so far.

     In this talk we report recent progress\cite{AICM} on the
incorporation of the effects due to the $\pi$~meson mass into the analytic
approach to QCD, and the study of the behavior of the developed running
coupling in spacelike and timelike infrared domains.

\section{Strong Running Coupling in Spacelike and Timelike Regions}

     The description of hadron dynamics in spacelike and timelike regions
has been the subject of many studies over a long period of time. The
strong interaction processes with the large spacelike momentum transfer
$q^2>0$ can be examined perturbatively in the framework of the RG method
(the metric with the signature $(-1,1,1,1)$ is used, so that positive
$q^2$ corresponds to a spacelike momentum transfer). But in order to
handle the processes in the timelike region ($s=-q^2>0$), one has first to
relate the perturbative results with the measured quantities.

     A substantial step towards the consistent description of the timelike
data was made in Refs.~\refcite{Adler}, \refcite{Rad82},
and~\refcite{KrPi82}. Namely, it has been argued that the dispersion
relation for the Adler $D$~function
\begin{equation}
\label{AdlerDisp}
D(q^2) = q^2 \int_{4\mpi^2}^{\infty} \frac{R(s)}{(s + q^2)^2}\, d s
\end{equation}
supplies a firm ground for comparing the perturbative results for~$D(q^2)$
with the experimental measurements of the $R(s)$--ratio of the $\epem$
annihilation into hadrons. Thus, the perturbative results for~$D(q^2)$ can
be continued into the timelike domain by making use of the relation
inverse to~(\ref{AdlerDisp}) (see also Refs.~\refcite{Rad82}
and~\refcite{GKL91})
\begin{equation}
\label{AdlerInv}
R(s) = \frac{1}{2 \pi i} \lim_{\varepsilon \to 0_{+}}
\int_{s + i \varepsilon}^{s - i \varepsilon}
D(-\zeta)\, \frac{d \zeta}{\zeta}.
\end{equation}

     The asymptotic ultraviolet behavior of the Adler $D$~function can be
computed by making use of the perturbation theory\cite{GKL91,SurSa91}
\begin{equation}
\label{AdlerPert}
D(q^2) = \Nc \sum_{f} Q_f^2 \left[1 + d(q^2)\right], \qquad
d(q^2) \simeq
d_1\left[\frac{\al{}{s}(q^2)}{\pi}\right]   +
d_2\left[\frac{\al{}{s}(q^2)}{\pi}\right]^2 + \cdots .
\end{equation}
Here $d_1 = 1$, $\,d_2 \simeq 1.9857 - 0.1153\,\nf$, $\,\nf$ is the number
of active quarks, $\Nc = 3$ is the number of colors, and $Q_f$~stands for
the charge of the $f$-th quark. Since the integral
transformation~(\ref{AdlerInv}) of Eq.~(\ref{AdlerPert}) has to be
performed every time when one deals with the timelike experimental data,
it is convenient to define\cite{MS97} the timelike effective charge
$\hal{}{}(s)$ in the same way, as $R(s)$ relates to~$D(q^2)$:
\begin{equation}
\label{TLviaSL}
\hal{}{}(s) = \frac{1}{2 \pi i} \lim_{\varepsilon \to 0_{+}}
\int_{s + i \varepsilon}^{s - i \varepsilon}
\al{}{}(- \zeta)\, \frac{d \zeta}{\zeta}.
\end{equation}
Here and further the strong running coupling in the spacelike and timelike
domains is denoted by $\al{}{}(q^2)$ and $\hal{}{}(s)$, respectively. The
inverse relation
\begin{equation}
\label{SLviaTL}
\alpha(q^2) = q^2 \int_{4 \mpi^2}^{\infty}
\frac{\hal{}{}(s)}{(s + q^2)^2}\, d s
\end{equation}
holds as well. Evidently, for the thorough description of the QCD
processes at low energies the pion mass cannot be neglected in
Eqs.~(\ref{AdlerDisp}) and~(\ref{SLviaTL}).

     For the self--consistency of the method described above, one has
first to bring the perturbative approximation for the Adler
$D$~function~(\ref{AdlerPert}) to conformity with the dispersion
relation~(\ref{AdlerDisp}). Indeed, Eq.~(\ref{AdlerDisp}) implies that in
the massless case ($\mpi=0$) $D(q^2)$ is the analytic function in the
complex $q^2$-plane with the only cut $-\infty < q^2 \le 0$ along the
negative semiaxis of real~$q^2$. However, the perturbative approximation
of $d(q^2)$ in Eq.~(\ref{AdlerPert}) violates this condition. Fortunately,
this discrepancy can be avoided in the framework of the analytic approach
to~QCD.

\section{Massless Analytic Effective Charge}

     As was mentioned in the Introduction, the dispersion relations
provide one with a certain nonperturbative information about a quantity in
hand, namely, with the definite analytic properties in the kinematic
variable. Recently it was argued\cite{ShSol} that in the massless case the
\KL spectral representation for the QCD invariant charge
\begin{equation}
\label{KLAn}
\alpha(q^2) = \int_{0}^{\infty}
\frac{\varrho(\sigma)}{\sigma + q^2} \, d \sigma
\end{equation}
holds. It is precisely this condition that is needed in order to bring the
perturbative approximation for the Adler $D$ function~(\ref{AdlerPert}) in
agreement with the dispersion relation~(\ref{AdlerDisp}) and to satisfy
the relations (\ref{TLviaSL}) and (\ref{SLviaTL}) in the limit of the
massless pion~$\mpi=0$.

     This section is devoted to a brief overview of one of the massless
models for the QCD analytic running coupling.\cite{PRD} The distinguishing
feature of this model is that the analyticity requirement is imposed on
the perturbative expansion of the $\beta$~function. This way of
incorporating the analyticity requirement into the RG formalism is
consistent with the general definition of the QCD invariant
charge.\cite{Review,MPLA2} The model\cite{PRD} has proved to be successful
in the description of the hadron dynamics of the both perturbative and
intrinsically nonperturbative nature.\cite{QCD03} The running
coupling\cite{PRD} has also been rederived in Ref.~\refcite{Schrempp}
proceeding from different assumptions.

     Thus, the \KL representation~(\ref{KLAn}) holds for the massless
analytic invariant charge (the relevant details can be found in
Refs.~\refcite{PRD} and~\refcite{Review})
\begin{equation}
\label{AICHLKL}
\al{(\ell)}{an}(q^2) = \fpb \int_{0}^{\infty}
\frac{\ro{\ell}(\sigma)}{\sigma + z}\, d \sigma,
\qquad z=\frac{q^2}{\Lambda^2}.
\end{equation}
In this equation $\ro{\ell}(\sigma)$ denotes the $\ell$-loop spectral
density
\begin{equation}
\label{SpDnsHL}
\ro{\ell}(\sigma) = \ro{1}(\sigma)
\exp\!\left[\pvint{0}{\infty}\!{\cal P}^{(\ell)}(\zeta)
\ln\!\left|1-\frac{\zeta}{\sigma}\right|
\frac{d \zeta}{\zeta}\right]\!
\left[\cos \psi^{(\ell)}(\sigma) +
\frac{\ln \sigma}{\pi} \sin \psi^{(\ell)}(\sigma)\right],
\end{equation}
where $\ro{1}(\sigma) = (1+\sigma^{-1})/(\ln^2 \sigma+\pi^2)$ stands for
the one-loop spectral density, $\psi^{(\ell)}(\sigma) = \pi
\int_{\sigma}^{\infty} {\cal P}^{(\ell)}(\zeta) \, \zeta^{-1} d\zeta$,
$\,{\cal P}^{(\ell)}(\sigma) = \Ro{\ell}{\sigma} - \Ro{1}{\sigma}$, and
\begin{equation}
\Ro{\ell}{\sigma} = \frac{1}{2 \pi i}
\lim_{\varepsilon \to 0_{+}} \sum_{j=0}^{\ell-1}
\frac{\beta_{j}}{(4\pi)^{j+1}}
\left\{\left[\al{(\ell)}{s}
(-\sigma - i \varepsilon)\right]^{j+1}-
\left[\al{(\ell)}{s}
(-\sigma + i \varepsilon)\right]^{j+1}\right\}.
\end{equation}
Here $\beta_0=11-2\nf/3,\; \beta_1=102-38\nf/3,\; \al{(\ell)}{s}(q^2)$ is
the $\ell$-loop perturbative running coupling. In the exponent in
Eq.~(\ref{SpDnsHL}) the principal value of the integral is assumed.

     The model~(\ref{AICHLKL}) shares all the advantages of the analytic
approach: it contains no unphysical singularities, displays good
higher-loop and scheme stability, and has no adjustable parameters. The
massless analytic effective charge~(\ref{AICHLKL}) incorporates the
ultraviolet asymptotic freedom with the infrared enhancement in a single
expression, that plays a crucial part when applying this model to the
study of quenched lattice simulation data.\cite{Schrempp} The detailed
analysis of the properties of the invariant charge~(\ref{AICHLKL}) and its
applications can be found in Refs.~\refcite{PRD}--\refcite{QCD03}.

     For the congruous description of the timelike hadron dynamics one has
to employ the continuation~(\ref{TLviaSL}) of the strong running coupling.
This leads to the following extension of the invariant
charge~(\ref{AICHLKL}) to the timelike domain\cite{PRD} (see also
Ref.~\refcite{MS97}):
\begin{equation}
\label{AICTL}
\hal{(\ell)}{an}(s) = \fpb\, \int_{w}^{\infty}\!
\ro{\ell}(\sigma)\,\frac{d \sigma}{\sigma},
\qquad w = \frac{s}{\Lambda^2},
\end{equation}
where $s=-q^2>0$, and the spectral density $\ro{\ell}(\sigma)$ is defined
in Eq.~(\ref{SpDnsHL}). The plots of the functions $\al{(1)}{an}(q^2)$ and
$\hal{(1)}{an}(s)$ are shown in Figure~1. Note the asymmetry between
these curves in the intermediate and low energy regions, which has to be
taken into account when one handles the experimental data (see
review\cite{Review} and references therein for the details).

     It is worth noting that the formulation of the analytic approach to
QCD is still essentially massless. The obtained results can be applied,
for example, to the study of the experimental data at high energies, the
pure gluodynamics, and the quenched lattice simulation data. However, for
a consistent description of the infrared hadron dynamics, the mass effects
have to be incorporated into the analytic approach to~QCD.

\section{Massive Analytic Effective Charge}

     The original dispersion relation for the Adler
$D$~function\cite{Adler}~(\ref{AdlerDisp}) with the nonvanishing mass of
the $\pi$~meson implies that $D(q^2)$ is the analytic function in the
complex $q^2$-plane with the only cut beginning at the two--pion threshold
$-\infty < q^2 \le -4\mpi^2$. However, the approximation~(\ref{AdlerPert})
violates this condition due to the spurious singularities of the
perturbative running coupling. This disagreement can be avoided by
imposing the analyticity requirement of the form $d(q^2, \mpi^2) = \int_{4
\mpi^2}^{\infty} \varkappa(\sigma)(\sigma + q^2)^{-1} d \sigma$ on the
function $d(q^2)$ in~Eq.~(\ref{AdlerPert}). Thus, the strong running
coupling itself has to satisfy the integral representation
\begin{equation}
\label{KLAnM}
\alpha(q^2, \mpi^2) = \int_{4 \mpi^2}^{\infty}
\frac{\varrho(\sigma)}{\sigma + q^2}\, d \sigma.
\end{equation}
Otherwise, one would encounter the contradiction between the dispersion
relation for the Adler $D$~function~(\ref{AdlerDisp}) and its
approximation~(\ref{AdlerPert}).

\begin{figure}[h]
\begin{center}
\begin{tabular}{lr}
\parbox[t]{59.25mm}{
\epsfig{file=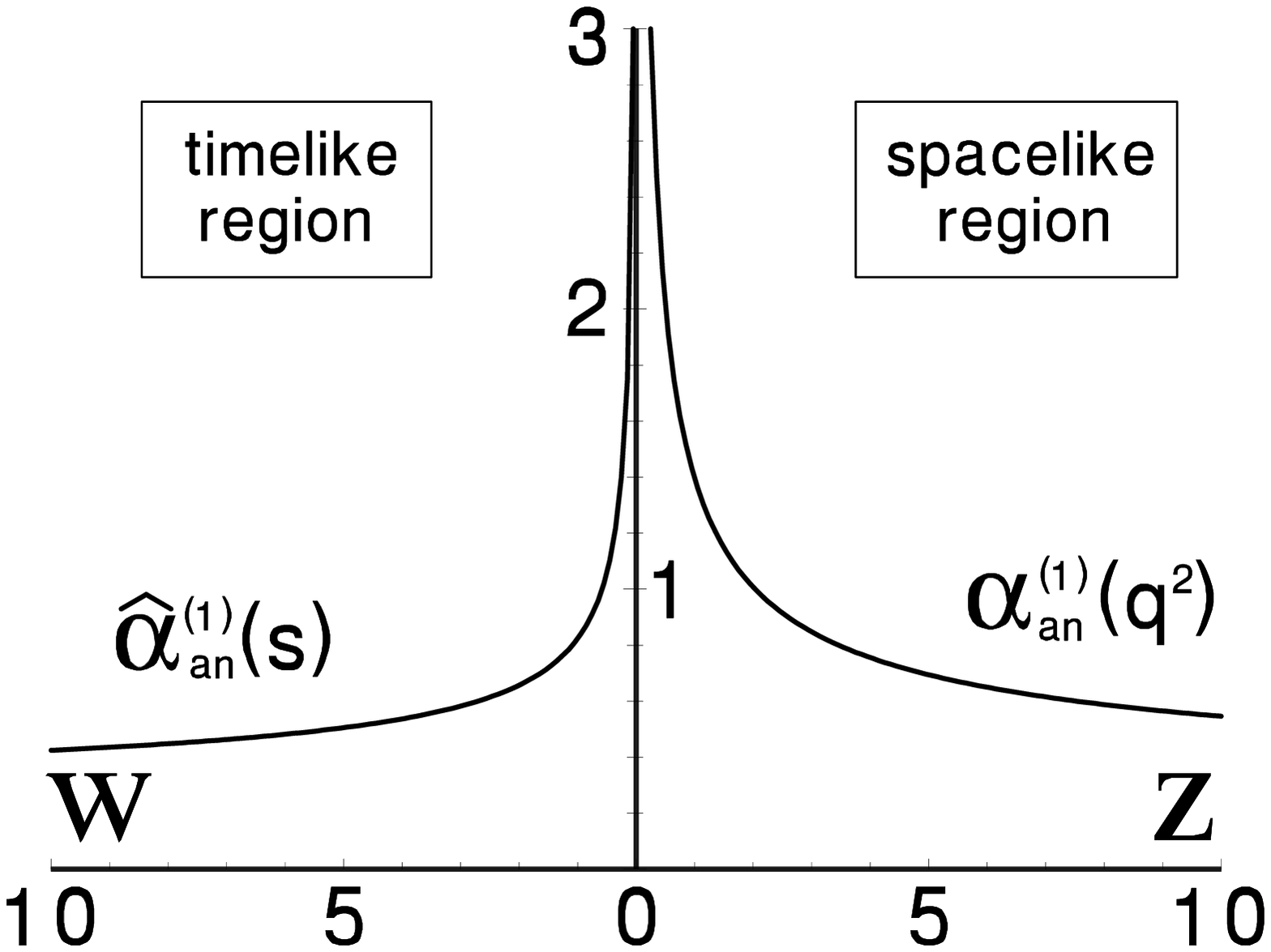, width=58mm}\hfill
\vskip1.5mm
\parbox[t]{58mm}{\footnotesize
Fig.~1. The one-loop massless analytic running coupling in spacelike
(Eq.~(\protect\ref{AICHLKL}), $q^2>0$) and timelike
(Eq.~(\protect\ref{AICTL}), $s=-q^2>0$) regions,
$z=-w=q^2/\Lambda^2$.}}\hfill
&
\parbox[t]{59.25mm}{
\hfill\epsfig{file=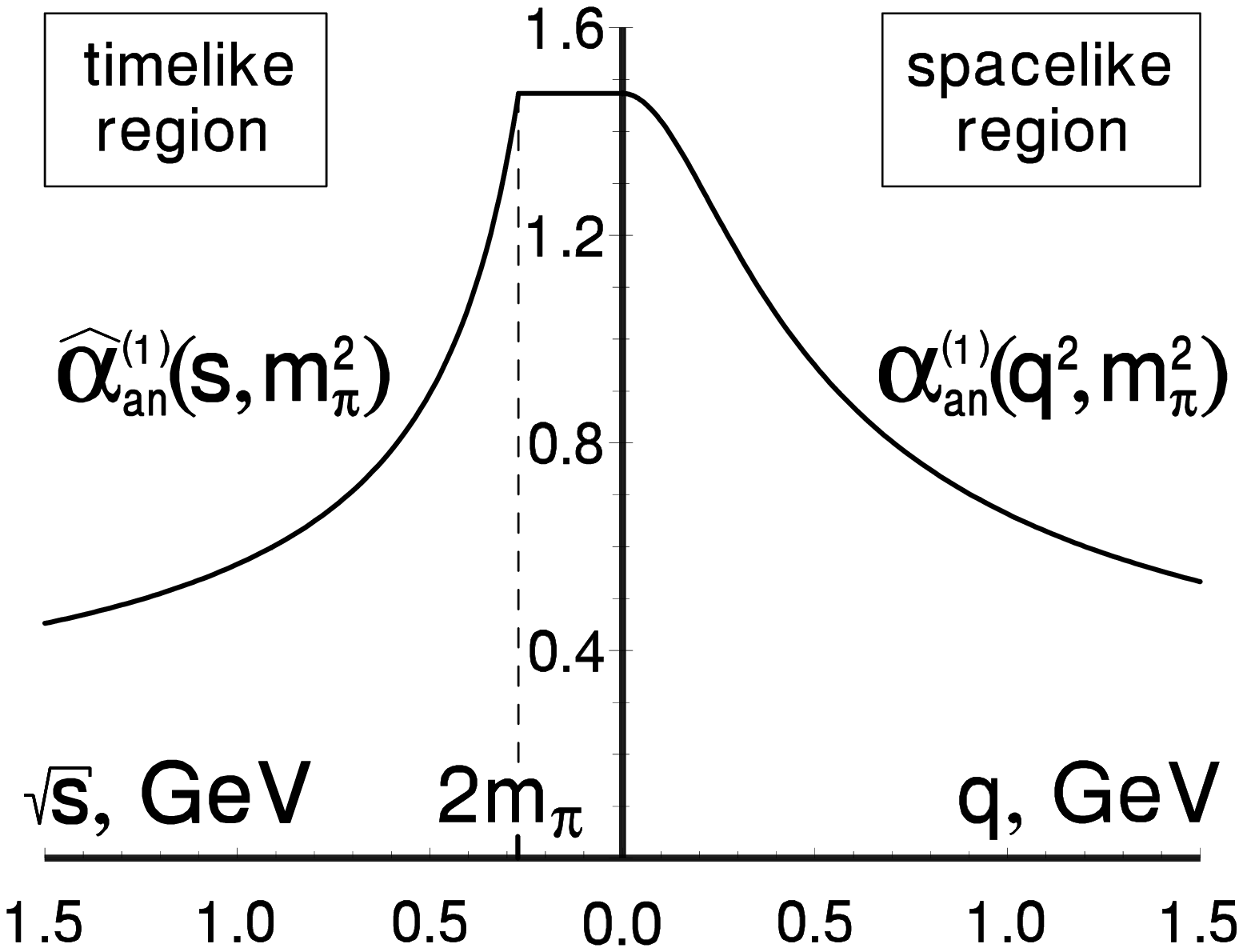, width=58mm}
\vskip1.5mm
\hfill\parbox[t]{58mm}{\footnotesize
Fig.~2. The one--loop massive analytic running coupling in spacelike
(Eq.~(\protect\ref{AICMHL})) and timelike (Eq.~(\protect\ref{AICMTL}))
domains. The value $\Lambda = 623\,$MeV and two active quarks are
assumed.}}
\end{tabular}
\end{center}
\end{figure}

     There are several ways to restore the spectral density
$\varrho(\sigma)$ in the integral representations~(\ref{KLAn})
and~(\ref{KLAnM}). Ultimately, this has led to different models for the
strong running coupling within the analytic approach to QCD (see
Refs.~\refcite{PRD}--\refcite{MPLA2},~\refcite{DV02}). The model for the
analytic invariant charge\cite{PRD} has proved to be successful in the
description of both perturbative and intrinsically nonperturbative strong
interaction processes.\cite{Review,QCD03} We will therefore adopt in this
work the spectral density~(\ref{SpDnsHL}).

     Thus, one arrives at the following integral representation for the
massive analytic effective charge
\begin{equation}
\label{AICMHL}
\al{(\ell)}{an}(q^2, \mpi^2) = \fpb\int_{\chi}^{\infty}
\frac{\ro{\ell}(\sigma)}{\sigma + z}\, d\sigma,
\qquad z=\frac{q^2}{\Lambda^2},
\end{equation}
where $\ro{\ell}(\sigma)$ denotes the $\ell$-loop spectral
density~(\ref{SpDnsHL}), and $\chi=4\mpi^2/\Lambda^2$. It turns out that
the nonvanishing mass of the $\pi$~meson drastically affects the infrared
behavior of the running coupling. Indeed, instead of the infrared
enhancement in the massless case~(\ref{AICHLKL}), one has here the
infrared finite limiting value for the effective charge~(\ref{AICMHL})
$\al{(\ell)}{0}(\chi) = 4\pi\beta_{0}^{-1}\!\int_{\chi}^{\infty}
\ro{\ell}(\sigma) \sigma^{-1} d\sigma$. The continuation~(\ref{TLviaSL})
of the running coupling~(\ref{AICMHL}) results in the following
representation for the timelike massive effective charge:
\begin{equation}
\label{AICMTL}
\hal{(\ell)}{an}(s, \mpi^2) = \fpb\int_{w}^{\infty}\!
\theta(\sigma - \chi)\, \ro{\ell}(\sigma)\, \frac{d \sigma}{\sigma},
\qquad w=\frac{s}{\Lambda^2},
\end{equation}
where $s = -q^2 \ge 0$, $\,\ro{\ell}(\sigma)$ is the $\ell$-loop spectral
density~(\ref{SpDnsHL}), and $\chi=4\mpi^2/\Lambda^2$.

     The effective charges~(\ref{AICMHL}) and~(\ref{AICMTL}) have a common
finite value $\al{(\ell)}{0}(\chi)$ in the infrared limit. Then, the
timelike coupling~(\ref{AICMTL}) has the ``plateau'' in the infrared
domain: $\hal{(\ell)}{an}(s,\mpi^2) = \al{(\ell)}{0}(\chi)$, when $0 \le
\sqrt{s} \le 2 \mpi$ (see Fig.~2). The massless (\ref{AICTL}) and massive
(\ref{AICMTL}) timelike couplings coincide for $\sqrt{s}>2\mpi$, since the
pion mass affects the coupling~(\ref{AICMTL}) only in the region
$0\le\sqrt{s}\le2\mpi$, where it stops running.

     Finally, the developed model has been applied\cite{AICM} to the study
of the experimental data on the inclusive $\tau$~lepton
decay,\cite{TauExp} that has given (at the one-loop level with two active
quarks) the value $\Lambda = (623 \pm 81)\,$MeV of the QCD scale
parameter.

\section*{Acknowledgments}

     A.N.~is grateful to I.L.~Solovtsov for valuable discussions and
comments. Support by grants RFBR (02-01-00601, 04-02-81025),
NS-2339.2003.2, and CICYT FPA20002-00612 is acknowledged.

\end{document}